\documentclass[prd,twocolumn,showpacs,preprintnumbers,%
aps,floatfix,amsmath,amssymb,amsfonts,letterpaper]{revtex4}

\usepackage{graphicx}
\usepackage{color}
\usepackage{rotating}
\usepackage{dcolumn}
\usepackage{subfigure}
\usepackage{tabularx,pifont,multirow}

\def\CP                {\ensuremath{C\!P}\xspace}
\newcommand{\rprime} {{\ensuremath{r^\prime}}\xspace}
\def\D       {\ensuremath{D}}
\def\DDstarm {\ensuremath{D^{(*)-}}}
\def\DDstarp {\ensuremath{D^{(*)+}}}
\def\resp#1{} 
\def\alphaDz{\ensuremath{\alpha_{\Dz}}\xspace}

\def\dstpdstm{\ensuremath{D^{*+}D^{*-}}\xspace}
\def\btodstpipm{\Bz\rightarrow \Dstarmp\pi^\pm}
\def\btodstdst{\ensuremath{\Bz \rightarrow \Dstarp\Dstarm}}

\def\mc{{Monte Carlo}\xspace}

\def\Brec{\ensuremath{B_{\rm rec}}}
\def\Btag{\ensuremath{B_{\rm tag}}}
\def\zrec{\ensuremath{z_{\rm rec}}}
\def\ztag{\ensuremath{z_{\rm tag}}}

\def\mrec{\ensuremath{m_{\rm rec}}\xspace}
\def\mep{\ensuremath{m_{\rm ep}}\xspace}
\def\Rperp{\ensuremath{R_\perp}\xspace}
\def\Sp{\ensuremath{S_+}\xspace}

\def\Cp{\ensuremath{C_+}\xspace}

\def\fisher{\ensuremath{F}}

\def\stag{\ensuremath{S_{\rm tag}}}

\def\dt {\ensuremath{\Delta t}}
\def\dtErr{\ensuremath{\sigma_{\dt}}}
\newcommand{\dm}[1]{\ensuremath{\Delta m_{{#1}}}}

\newcommand{\ModeO}{\ensuremath{K\pi}}
\newcommand{\ModeT}{\ensuremath{K\pi\piz}}
\newcommand{\ModeTh}{\ensuremath{K\pi\pi\pi}}
\newcommand{\ModeF}{\ensuremath{\KS\pi\pi}}

\def\pdf{PDF}
\def\pisoft{\ensuremath{\pi_s}}
\def\fBB{\ensuremath{f_{\BB}}}

\def\comb{{\rm comb}}
\def\P{{\cal P}}  
\def\T{{\cal T}}  
\def\M{{\cal M}}  
\def\F{{\cal F}}  
\def\R{{\cal R}}  
\def\true{{\rm true}}
\def\dttrue{\ensuremath{\dt_\true}}

\def\bea{\begin{eqnarray}}
\def\eea{\end{eqnarray}}
\def\Sbkg {\ensuremath{S_{\rm comb}\xspace}} 
\def\Cbkg {\ensuremath{C_{\rm comb}\xspace}} 
\def\G{\ensuremath{{\cal G}}\xspace}  

\newcommand\Tp{\rule{0pt}{2.6ex}}	
\newcommand\Bt{\rule[-1.2ex]{0pt}{0pt}}	
\def\Bztodstadsta {\ensuremath{\Bz \to D^{(*)+} D^{(*)-}}\xspace}
\def\babar{\mbox{\slshape B\kern-0.1em{\smaller A}\kern-0.1em
    B\kern-0.1em{\smaller A\kern-0.2em R}}}

\input pubboard/babarsym

\begin{document}

\preprint{SLAC-PUB-15205}
\preprint{{\babar}-PUB-12/022}
\preprint{arXiv:1208.1282 [hep-ex]}

\title {\boldmath Measurement of the Time-Dependent \CP Asymmetry of Partially Reconstructed \Bztodstdst\ Decays}

%
\author{J.~P.~Lees}
\author{V.~Poireau}
\author{V.~Tisserand}
\affiliation{Laboratoire d'Annecy-le-Vieux de Physique des Particules (LAPP), Universit\'e de Savoie, CNRS/IN2P3,  F-74941 Annecy-Le-Vieux, France}
\author{J.~Garra~Tico}
\author{E.~Grauges}
\affiliation{Universitat de Barcelona, Facultat de Fisica, Departament ECM, E-08028 Barcelona, Spain }
\author{A.~Palano$^{ab}$ }
\affiliation{INFN Sezione di Bari$^{a}$; Dipartimento di Fisica, Universit\`a di Bari$^{b}$, I-70126 Bari, Italy }
\author{G.~Eigen}
\author{B.~Stugu}
\affiliation{University of Bergen, Institute of Physics, N-5007 Bergen, Norway }
\author{D.~N.~Brown}
\author{L.~T.~Kerth}
\author{Yu.~G.~Kolomensky}
\author{G.~Lynch}
\affiliation{Lawrence Berkeley National Laboratory and University of California, Berkeley, California 94720, USA }
% ...
\collaboration{The \babar\ Collaboration}
\noaffiliation

\date{31 October, 2012}

\begin{abstract}
We present a new measurement of the time-dependent \CP asymmetry of
\Bztodstdst\ decays using ($471\pm5$) million \BB\ pairs
collected with the \babar\ detector at the PEP-II \B Factory at the SLAC
National Accelerator Laboratory. 
Using the technique of partial reconstruction, we measure the time-dependent \CP asymmetry parameters $S=-0.34 \pm 0.12 \pm 0.05$ and $C=+0.15 \pm 0.09 \pm 0.04$.
Using the value for the \CP-odd fraction $R_\perp=0.158 \pm
0.028\pm 0.006$, previously measured by \babar\ 
with fully reconstructed \Bztodstdst events, we extract the \CP-even
components $\Sp=-0.49 \pm 0.18 \pm 0.07 \pm 0.04$ and $\Cp=+0.15 \pm 0.09 \pm 0.04$.
In each case, the first uncertainty is statistical and the second is
systematic; the third uncertainty on \Sp\ is the contribution from the
uncertainty on $R_\perp$.
The measured value of the \CP-even component $S_+$ is consistent
 with the value of $\sin2\beta$ measured in
$b\to~(\ccbar)s$ transitions, and with the Standard Model expectation of small penguin contributions.

\end{abstract}

\maketitle
\vfill
\centerline{(Submitted to \pr{D})}

\section{Introduction}
\label{sec:Intro}
%
%

In the Standard Model (SM), \CP violation arises from an irreducible complex phase
in the $3\times 3$ quark mixing matrix $V$ known as the 
Ca\-bib\-bo-Ko\-ba\-ya\-shi-Mas\-ka\-wa (CKM) matrix\,\cite{Cabibbo:1963yz,Kobayashi:1973fv}.
Unitarity of the CKM matrix requires that the relation
$V_{ud}V^*_{ub}+V_{cd}V^*_{cb}+V_{td}V^*_{tb}=0$, 
which defines the unitarity triangle, be satisfied.
The aim of the \B Factories is  to test the unitarity of the CKM matrix
by the precise measurement of the angles and sides of the above
triangle, whose nonvanishing area indicates violation of \CP symmetry.

Both the \babar~and Belle collaborations have measured the \CP
parameter \stwob, where the angle $\beta$ is defined as 
$\beta\equiv\text{arg}\left[-V_{cd}V_{cb}^*/V_{td}V_{tb}^*\right]$. 
The most accurate measurements of \stwob\,\cite{Aubert:2009ac,Chen:2007ur,Sahoo} 
use the $b\rightarrow(c\cbar)s$ transition, in which \Bz's decay to charmonium final states. 
Measurement of $b\rightarrow c\cbar d$ transitions such as \Bz\to\DDstarp\DDstarm
should yield the same value of \stwob to the extent that the contributions
from penguin processes may be neglected.

\begin{figure}[bt]
\vspace{-10pt}
\subfigure[\label{fig:tree}~Tree]{\includegraphics[width=0.4\columnwidth]{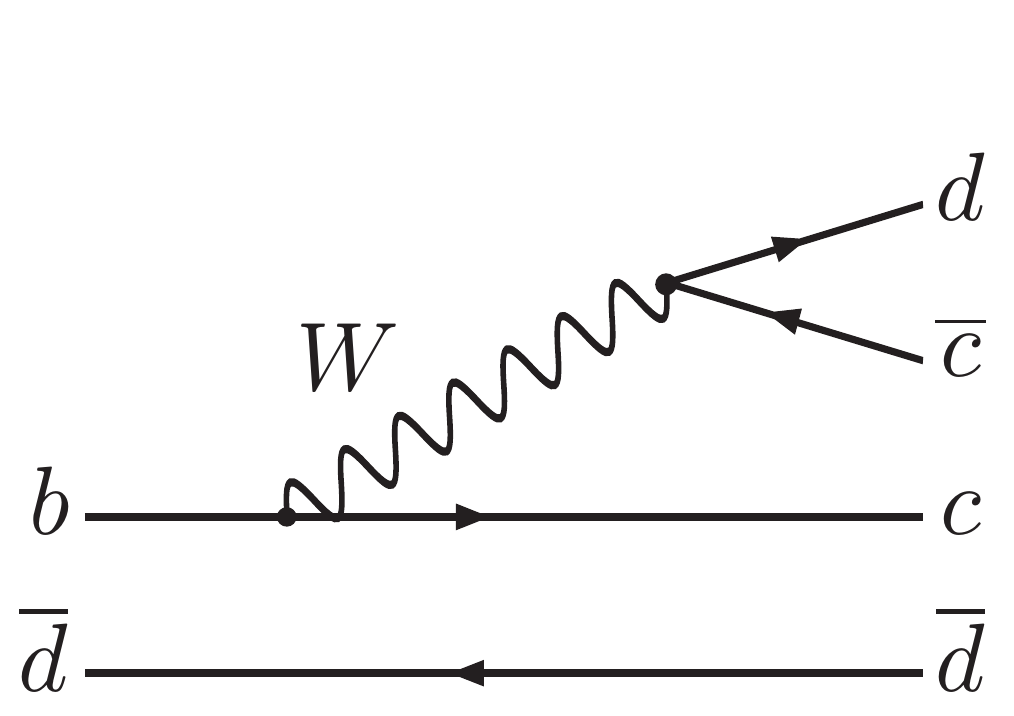}}\quad
\subfigure[\label{fig:penguin}~Penguin]{\includegraphics[width=0.4\columnwidth]{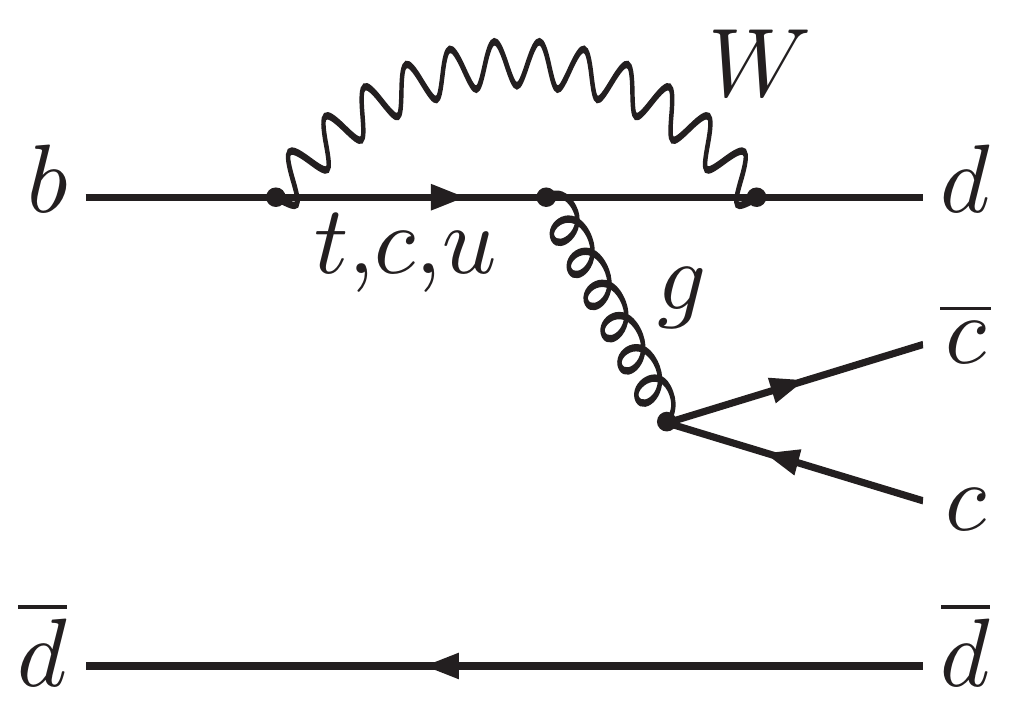}}
\vspace{-10pt}
\caption{\label{fig:diags}Leading and sub-leading order Feynman graphs for the \Bzb\to\DDstarp\DDstarm decays.}
\end{figure}

The leading and sub-leading order Feynman diagrams contributing to \Bztodstadsta decays are
shown in Fig.~\ref{fig:diags}.
The effect of neglecting the penguin amplitude has been estimated in models 
based on factorization and heavy quark symmetry, and the corrections 
are found to be a few percent\,\cite{Xing:1998ca,Xing:1999yx}.
Loops involving non-SM particles (for example, charged Higgs or SUSY particles)
could increase the contribution from penguin diagrams and introduce additional phases.

In $\Upsilon(4{\rm S})\to \Bz\Bzb$ events the time-dependent decay rate for 
\Bztodstdst\ is given by
\begin{eqnarray}
\label{eq:pure-dt-pdf-B}
\P^{\stag}_\eta(\dt)
&=& {e^{-|\dt|/\tau_b} \over 4\tau_b} \cdot       
\left[ 1 + \stag\, S_\eta\sin(\dm{d} \dt) \right.  \nonumber\\  
& +& \left. \stag\, C \cos(\dm{d} \dt) \right],
\end{eqnarray}
where $\tau_b$ is the $\Bz$ lifetime averaged over the two mass eigenstates,
\dm{d}\ is the $\Bz \Bzb$ mixing frequency, and $\dt$
is the time interval between the \Bztodstdst\ decay ($\Brec$)
and the decay of the other $B$ ($\Btag$) in the event. 
The parameter $\stag=+1~(-1)$  in Eq.~(\ref{eq:pure-dt-pdf-B})
indicates the flavor of the $\Btag$ as a $\Bz$ ($\Bzb$),
while $\eta = \pm 1$  indicates the \CP eigenvalue of the \Bztodstdst\ final state.
The parameters $C$ and $S_\eta$ are given by
\begin{equation}
  C = \frac{1-|\lambda|^2}{1+|\lambda|^2}; \quad
  S_\eta = -\eta\frac{2\Im m(\lambda)}{1+|\lambda|^2};\quad
  \lambda =\frac{q}{p}\,\frac{\overline A}{A},
\label{eq:CandS}
\end{equation}
\noindent where 
$A~(\overline A)$ is the matrix element
of the \Bz (\Bzb) decay and $p$ and $q$ are the coefficients appearing in the expression
of the physical mass eigenstates $B_L$, $B_H$ in terms of the flavour
eigenstates $B$, $\Bbar$:
\begin{eqnarray}
|B_L\rangle&=p|B\rangle+q|\Bbar\rangle\nonumber\\
|B_H\rangle&=p|B\rangle-q|\Bbar\rangle\nonumber.
\end{eqnarray}
Since the \Bztodstdst is the decay of a scalar to two vector mesons,
the final state is a mixture of \CP eigenstates. The \CP-odd and \CP-even fractions
have been previously measured from the angular analysis of completely reconstructed events\,\cite{Aubert:2009rx,Vervink:2009sy}.

A large deviation of the measured parameter $S_\eta$ in Eq.\,\ref{eq:CandS}
from the value of \stwob measured in $b\rightarrow(c\cbar)s$ transitions
or a non-zero value of direct \CP violation\,\cite{Grossman:1996ke,Gronau:2008ed,Zwicky:2007vv}
would be strong evidence of new physics.

Both the \babar\,\cite{Aubert:2009rx} and Belle\,\cite{Vervink:2009sy} collaborations
have studied the \CP asymmetries of \Bztodstdst decays 
using fully reconstructed events.
In this article we report a new measurement based on
the technique of partial reconstruction, which allows us to gain a
factor of $\simeq 5$ in the number of selected signal events 
with respect to the most recent \babar\ full reconstruction analysis in\,\cite{Aubert:2009rx}. 
This result is complementary to the latter measurement,
because the statistics used are largely independent of each other.

\section{The \babar\ Detector and Dataset}
\label{sec:DetDataSet}
%
%
The data sample used in this analysis has been collected with the \babar\
detector\,\cite{Aubert:2001tu} operating at the \pep2\ asymmetric-energy \BF\ located at
the SLAC National Accelerator Laboratory. 
We have analyzed the full \babar\ data set collected at the the \FourS mass peak, $\sqrt{s}=10.58 \gev$,
corresponding to an integrated luminosity of 429.0 \invfb. In addition,
we have used 44.8 \invfb of data taken off-resonance to evaluate the background
from events $\epem \rightarrow \qqbar$, where $q$ represents a $u,d,s$ or $c$ quark (``continuum'').
To study backgrounds and validate the analysis procedure, we use a {\tt GEANT4}-based\,\cite{Agostinelli:2002hh} \mc(MC) 
simulation in which coherent \BB\ production is simulated using the package {\tt EvtGen}\,\cite{LangeRyd:1998}.

The asymmetric energies of the \pep2\ beams are an ideal
environment to study time-dependent \CP phenomena in the $\Bz$-$\Bzb$
system. The boost of the \FourS in the laboratory frame by $\beta\gamma=0.56$  
increases the separation between the vertices of the two $B$ meson daughters, 
allowing their precise measurement.

The \babar\ detector is described in detail in Ref.\,\cite{Aubert:2001tu}.
We give here only a brief description of the main components and their use in
this analysis. Tracking is provided by a five-layer silicon vertex detector (SVT) and a drift chamber (DCH).
The SVT provides precise position measurements
 close to the interaction region that are used in vertex
 reconstruction and low-momentum track reconstruction.
The DCH provides excellent momentum measurement of charged particles. 

Particle identification (PID) of kaons and pions is obtained
from ionization losses in the SVT and DCH and from
measurements of photons produced in a ring-imaging Cherenkov light detector (DIRC).  
A CsI(Tl) crystal-based electromagnetic calorimeter (EMC) enables measurement of photon energies and
electron identification. These systems all operate inside a
1.5 T superconducting solenoid, whose iron flux return is instrumented for muon detection,
initially with resistive plate chambers and more recently with limited streamer tubes\,\cite{LST}.

\section{Analysis Method}
\label{sec:AnalMethod}
%
%
\subsection{Partial Reconstruction}
\label{sec:PartRec}
In the partial reconstruction of a \Bztodstdst candidate, we reconstruct fully only one of
the two \Dstarpm mesons in the decay chain \Dstar\to $D^0\pi$~\cite{cconjimplied},
by identifying \Dz candidates in one of four final states: \ModeO, \ModeT, \ModeTh, \ModeF.
The vertexing algorithm fits the two-step decay tree simultaneously, 
correctly calculating correlations among all candidates.
In the first three \Dz decay modes, assumed to represent Cabibbo-favored decays,
charged kaon tracks are selected using PID information from the DIRC, SVT and DCH.
In the last decay mode, \KS candidates are selected by constraining pairs of
oppositely charged tracks to a common vertex.

\begin{figure}[htbp]
    \vspace{-35pt}
\begin{center}
\begin{tabular}{cc}
        \includegraphics[width=0.5\textwidth]{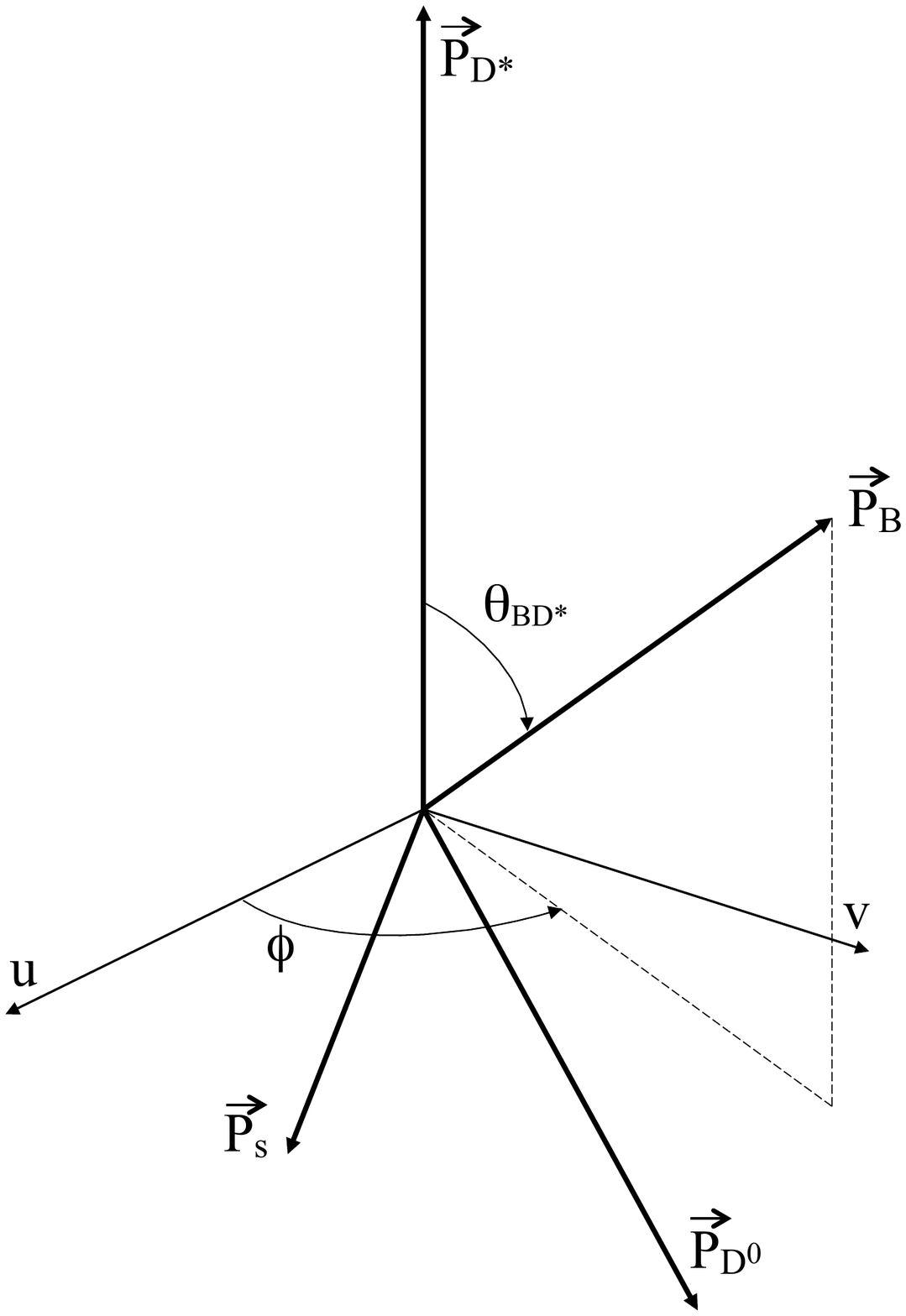}
\end{tabular}
\end{center}
    \vspace{-35pt}
\caption{Momenta and angles in the \FourS center of mass frame used in partial reconstruction.
The orthogonal axes $u$ and $v$ are normal to the
momentum $\vec{p}_{\Dstar}$ of the reconstructed \Dstar, and $u$ lies in the plane
defined by the momenta of the \Dstar and slow pion, $ \vec{p}_{\Dstar}$ and $\vec{p}_s$. 
The angle $\phi$ is measured in the $u-v$ plane.}
\label{fig:dstdst-sketch}
\end{figure}

Since the kinetic energy available in the decay \Dstar\to\Dz$\pi$ is small,
we combine one reconstructed \Dstarpm with an oppositely charged low-momentum pion,
assumed to originate from the decay of the unreconstructed \Dstarmp,
and evaluate the mass \mrec\ of the recoiling \Dz meson by using the
momenta of the two particles.
For signal events \mrec peaks at the nominal \Dz mass\,\cite{ref:PDG2010}
with an r.m.s.~width of about 3~\mevcc, while for background events no such peak is visible.
Thus, \mrec is the primary variable to discriminate signal from background.
The calculation of \mrec proceeds as follows 
(refer to Fig.~\ref{fig:dstdst-sketch} for definitions of the various momenta and angles that we use).

The cosine of the angle between the momenta in the \FourS center of mass (CM)
frame of the \B and the reconstructed \Dstar 
is readily computed:
\begin{equation}
\cos\theta_{B\Dstar} =
        {- M_{\Bz}^2 + E_{\rm CM} E_{\Dstar}
                        \over
         2 p_B |\vec p_{\Dstar}|
        },
\label{eq:cosTheta}
\end{equation}
where all particle masses are set to their nominal values\,\cite{ref:PDG2010},
$E_{\Dstar}$ and $\vec p_{\Dstar}$
are the measured energy and momentum of the reconstructed \Dstar in the \Y4S CM frame,
$E_{\rm CM}$/2 is the energy of each beam in the CM frame, and
$p_B = \sqrt{E_{\rm CM}^2/4 - M_{\Bz}^2}$ is the $B$-meson CM momentum.
Events are required to be in the physical region  $|\cos\theta_{B\Dstar}|<1$.

Given $\cos\theta_{B\Dstar}$ and the measured momenta of the \Dstar and 
oppositely charged low-momentum pion, \pisoft,
the \B four-momentum can be calculated up to
an unknown azimuthal angle $\phi$ around ${\vec p}_{\Dstar}$. 
For any chosen value of $\phi$, conservation laws determine the unreconstructed \Dz\ 
four-momentum $q_D(\phi)$, and one can thus compute the corresponding $\phi$-dependent 
invariant mass $m(\phi) \equiv \sqrt{|q_D(\phi)|^2}$.
The value of $\phi$ is not constrained by kinematics and may be chosen arbitrarily,
to the extent that the shape of the resulting $m(\phi)$ distribution may still be
described by the type of functions used in our fits.
We have chosen the value for which $\cos\phi=0.62$, which is the median of the 
corresponding \mc distribution for signal events
obtained using generated momenta, and define the recoiling \Dz mass
$\mrec\equiv m(\cos\phi\!=\!0.62)$.
We use the same convention to obtain the direction of the unreconstructed \Dz meson.

\subsection{Backgrounds and Event Selection}
\label{sec:cuts}

Backgrounds to the \Bztodstdst  process include 
the following:
\begin{itemize}
\item Combinatorial \BB background, defined as decays other than \Bztodstdst,
for which the \mrec distribution is approximately flat.
\item Peaking $\BB$ background, defined as decays other than \Bztodstdst, 
in which the $\mrec$ distribution peaks in the signal region. 
It will be shown later that the contribution from this background is negligible.
\item  Background from non-$b\bbar$ events.
\end{itemize}

Combinatorial \BB background events are reduced by the following requirements:
For the \ModeF\ mode, we require the invariant mass of the pion pair
to be within 25 \mevcc of the \KS mass\,\cite{ref:PDG2010}.
The corresponding vertex must be separated by more than 3 mm from the beam axis. 
For the \ModeT\ mode, \piz candidates are formed from pairs of photons
detected in the EMC, with energies greater than 40\mev, for which the invariant mass
differs by less than 20\mevcc from the nominal \piz mass\,\cite{ref:PDG2010}. 
The reconstructed \Dz mass must be equal to the nominal one\,\cite{ref:PDG2010} within 2 or 2.5
standard deviations, depending on the \Dz reconstruction mode.
The momenta in the $\Upsilon(4{\rm S})$ CM frame of the reconstructed \Dstar and \pisoft~from the missing \Dz 
must be, respectively, in the range 1.3--2.1\gevc and smaller than 0.6\gevc.
The difference $\Delta M =|M_{\Dstar}-M_{\Dz}-M_{\pi}|$ must be equal to the nominal\,\cite{ref:PDG2010} value 
within 1 or 1.5\mevcc, according to the presence or absence of DCH hits in the 
pion track appearing in the reconstructed decay \Dstar\to\Dz$\pi$.
The probability of the vertex fits must be greater than $10^{-2}$, for both the \Dz and the \Dstar reconstruction.

The requirement on the \Dz\ vertex fit probability introduces a small but measurable bias 
toward lower values of the $B$ lifetime. 
Due to partial reconstruction, the tracks used to make the \Dz\ vertex may originate from
the same or different $B$ mesons. In the latter case, since not all tracks are from
the same point in space, the $\chi^2$ of the vertex fit tends to be bigger. This 
effect worsens with increasing distance between the two $B$ decay vertices, 
causing vertices further apart to be rejected more frequently. 
We have verified this on signal \mc\ events, for which we have
measured a lifetime lower than the generated value. 
Consequently, for the signal \dt\ probability distribution functions (\pdf's) 
we use the value of $\tau_b$ fitted to signal \mc.

In events passing this selection we find more than one candidate decay chain in about 25\% of the cases,
usually differing only in the slow pion \pisoft, but sometimes in the components of the
reconstructed \Dstar. When this happens, we choose one candidate chain, based respectively
on the largest number of DCH hits in the \pisoft, or according to a $\chi^2$ based on the 
reconstructed \Dz mass and $\Delta M$ quantity above.
For signal \mc, the probability for this candidate chain
to be the correct one is 0.95.

The main suppression of continuum background is obtained by requiring
that the ratio $R_2$ of the $2^{nd}$ to the $0^{th}$ Fox-Wolfram moment\,\cite{ref:R2}, 
computed using all charged particles and EMC clusters not matched to tracks, be less than 0.3.

\subsection{Fisher Discriminant}
\label{sec:fisher}
To further reduce continuum background, we combine several event-shape 
variables into a Fisher discriminant\,\cite{ref:fisher} $F$.
Discriminating power originates from the observation that $\qqbar$ events tend to
be jet-like, whereas $\BB$ events have a more spherical energy
distribution. Rather than applying requirements on $F$,
we use the corresponding distribution in the fits described in Sec.\,\ref{sec:pdf}

Our Fisher discriminant is a linear combination of variables
chosen, according to \mc studies, to maximize the separation 
between \BB\ and continuum events. The first nine variables
describe the energy flow inside nine concentric cones centered
around the direction of the reconstructed \Dstarpm. In addition, we use 
the momenta of the charged and the neutral particle closest to the cone axis,
the polar angles in the CM of the reconstructed \Dstar momentum and the thrust axis $T$
for charged tracks in the \Btag\, vertex (see next paragraph),
the angle between the reconstructed \Dstar momentum and $T$, and
the sum $S=\Sigma_i p_i\times P_2(\cos\theta_i)$ over the \Btag\, charged tracks, 
in which $p_i$ is momentum, $P_2$ is the 2$^{nd}$ Legendre polynomial of argument $\cos\theta_i$, 
and $\theta_i$ is the angle between track $i$ at the origin and $T$.

\subsection{Flavor Tagging and Decay Time Measurement}
\label{sec:flatAndDt}
For this analysis, two measurements are needed: the difference $\dt$ between
the proper decay times of the partially reconstructed \B meson and
the other \B meson in the event, and the flavor of the latter. 

The flavor tagging algorithm is based on tracks identified as electrons, muons or kaons.
The electron and muon tags contribute equally to the total sample and, since these events  
are kinematically almost indistinguishable and have very similar effective tagging efficiency, 
we treat them as one homogeneous ``lepton'' sample. 

The tagging tracks must be chosen among those not used in \Brec\ reconstruction and
must originate from within 4 mm (3 cm) of the interaction point in 
the transverse (longitudinal) view.
The momentum of the lepton candidates is required to be greater than 1.1
GeV/c in order to reject most leptons from charmed meson decays. 
If one or more lepton candidates are qualified, the tag flavor is assigned based on the
charge of the lepton with the highest center-of-mass momentum. 
If two or more qualified kaons are present,
the event is used only if the flavor is unambiguous.
If both a lepton and a kaon tag are available, the lepton tag is used.

The time difference $\dt$ is calculated using $\dt = \Delta z / \gamma\beta c$, where
$\Delta z=\zrec -\ztag$ is the difference between the $z$-coordinates of
the partially reconstructed \Brec\ and \Btag\ vertices and the boost parameters
are calculated using the measured beam energies.
The uncertainty $\dtErr$ on $\dt$ is calculated from the results
of the $\zrec$ and $\ztag$ vertex fits. We require $|\dt| < 20 \ps$ and 
 $\dtErr < 2.5 \ps$. 

We define the \Brec\ vertex as the decay point of the fully reconstructed \Dstarpm.
The \pisoft~track from the other \Dstarpm is not used, since it undergoes significant
multiple Coulomb scattering and hence does not improve 
the $\zrec$ measurement resolution.

The \Btag\ vertex reconstruction depends on the tagging category.
For kaon-tagged events, we obtain $\ztag$ from a beam spot constrained
vertex fit of all charged tracks in the event, excluding those from the \Brec\ meson,
and excluding also tracks within 1~rad of the unreconstructed \Dz momentum in the CM frame, 
which presumably originate from the \Dz decay. We require the probability
of this fit to be greater than $10^{-2}$.
For lepton-tagged events, we use the lepton track parameters and errors,
and the measured beam spot position and size in the plane perpendicular to the
beams (the $x$-$y$ plane). We find the position of the point in space
for which the sum of the $\chi^2$ contributions from the lepton track
and the beam spot is minimum. The $z$ coordinate of this point
is taken as $\ztag$.

The beam spot is measured on a run-by-run basis using 2-prong events (Bhabha and \mumu),
and has an r.m.s.~size of approximately 120~$\mu$m in the horizontal
dimension ($x$), 5~$\mu$m in the vertical dimension ($y$), and
8.5~mm along the beam direction ($z$). The average \B meson flight distance in the $x$-$y$ plane 
is 30~$\mu$m. To account for the \B flight distance in the beam spot constrained vertex fit, 
30~$\mu$m are added in quadrature to the effective $x$ and $y$ sizes.
\subsection{Probability Distribution Functions}
\label{sec:pdf}
%
%
%
We use two \pdf's, $P_{{\rm on}}$ for on-resonance,
and $P_{{\rm off}}$ for off-resonance data.
The former depends on the variables
\mrec, \fisher, \dt, \dtErr, 
\stag,
and is given by the sum of the \pdf's for the different event types described above:
\begin{eqnarray}
\P_{{\rm on}} = \fBB \, [ f_{\rm sig}\P_{\rm sig} + (1 - f_{\rm sig}) \P_\comb
                 ] +  (1 - \fBB ) \P_{\qqbar}
\label{eq:main-pdf}
\end{eqnarray}
\noindent where $\P_{\rm sig}$,  $\P_\comb$, and $\P_{\qqbar}$ are respectively the \pdf's for signal events, 
for combinatorial background from \BB, and for continuum. Moreover, \fBB\ is the fraction of \BB\ events in our sample,
and $f_{\rm sig}$ is the fraction of signal events in \BB\ events.
The \pdf\ for off-resonance data, $P_{{\rm off}}$, is reduced to just one component, $\P_{\qqbar}$, 
as the off-peak sample contains only continuum events. 

According to \mc, the distributions of \BzBzb\ and \BpBm\ combinatorial background events are very similar
and can be described well by the same \pdf.

We do not consider the fraction of \BB\ events a free parameter, but fix it to $f_{\BB}=1-f_{\qqbar}$, 
where $f_{\qqbar}$ is the fraction of continuum events in the on-peak sample and is defined by
\begin{equation}
f_{\qqbar}=\frac{ N_{{\rm off-peak}} }{ N_{{\rm on-peak}} }\frac{ {\cal L}_{{\rm on-peak}} }{ {\cal L}_{{\rm off-peak}} } ,
\label{eq:fbb}
\end{equation}
where $N$'s are the number of events left by our selection in the on- and off-peak samples 
and ${\cal L}$'s are the integrated on- and off-peak luminosities.

Each of the $\P_i$ ($i=$ sig, comb, \qqbar) can be expressed as the product of three 
one dimensional \pdf's:
\begin{eqnarray}\label{eq:prod-pdf}
\P_i(\mrec, F, \dt, \dtErr, \stag)  = \\
\M_i(\mrec) \, \F_i(F) \, \T'_i(\dt, \dtErr, \stag),\nonumber
\end{eqnarray}
that are the probability distributions of the recoil \Dz\ mass $\M_i(\mrec)$, 
the Fisher discriminant function $\F_i(F)$, and the decay time difference function $\T'_i(\dt, \dtErr, \stag)$.
This follows from extensive \mc\ studies showing that the correlations
among these variables are negligible. 

\subsubsection{$\M(\mrec)$ and $\F(F)$ \pdf's}



The \mrec\ distribution of all sample components can be well modelled in the lower region of the spectrum with a
so called ``Argus function''\,\cite{argus}:
\begin{equation}
\label{eq:argusPdf}
{\cal A}(\mrec) = \mrec\sqrt{1-(\mrec/m_{\rm ep})^2} \cdot e^{c\cdot\mrec /m_{\rm ep}},
\end{equation}
where $m_{\rm ep}$ is the kinematic endpoint ($\mrec \leq m_{\rm ep}$) and $c$ is a free parameter describing the slope.
This function alone, however, is not sufficient to account for the abrupt fall of the \mrec\ spectrum near the kinematic endpoint.
For the signal sample we model this shoulder with 
an asymmetric error function with widths $\sigma_l$ and $\sigma_r$, tapered
off at low \mrec\ by an exponential factor with decay constant $a$:
\begin{equation}
\begin{array}{l}
\label{eq:bifurPdf}
{\cal E}(\mrec) =\\
\Biggl\{ 
\begin{matrix}
  e^{\mrec/a}[1-{\rm erf}(\mrec-\mep)/({\sqrt{2}\sigma_l})], &\mrec <\mep\\
  e^{\mrec/a}[1-{\rm erf}(\mrec-\mep)/({\sqrt{2}\sigma_r})], &\mrec >\mep.
\end{matrix} 
\end{array}
\nonumber
\end{equation}
Thus, we describe the signal \mrec distribution with a combination of three functions:
a Gaussian $G$ having average $m_G$ and standard deviation $\sigma_G$ for the  well reconstructed peaking component; 
an Argus function, mainly for events in which the right \Dstar\ is combined 
with a low-momentum pion from another decay chain; and the ${\cal E}$ function:
\begin{equation}
\begin{array}{lll}
\M_{{\rm sig}}(\mrec) & = &  f_{{\rm sig}}^{\cal A} \cdot {\cal A}(\mrec)\,+\cr
& + &(1 -f_{{\rm sig}}^{\cal A})\cdot\Bigl[f^{G}\cdot G(\mrec)\,\cr 
& + &\, (1 - f^{G})\cdot{\cal E}(\mrec)\Bigr].
\end{array}
\label{eq:mrecSigPdf}
\end{equation}
In Eq.\,\ref{eq:mrecSigPdf} $f_{{\rm sig}}^{\cal A}$ is the fraction of events described by the Argus component
and $f^{G}$ is the fraction of events in the Gaussian peak relative to the non-Argus
component.


For the background, both combinatorial and continuum, we set the fraction of the Gaussian component to zero, 
and model the distribution at the endpoint with a simple error function of width $\sigma$.
However, for the case of combinatorial background in kaon-tagged events,
we find that two different Argus components (${\cal A}_1$ and ${\cal A}_2$) are needed
to correctly describe the 
entire 
reconstructed mass spectrum. We thus define two \pdf's according to:
\begin{equation}
\begin{array}{lll}
\M_{\rm comb}(\mrec) & = & f_{\rm comb}^{\rm erf} \cdot {\rm erf}(\mrec-m_{{\rm ep}}; \sigma_{\rm comb})\, +\cr
& + &\,(1 -f_{\rm comb}^{\rm erf})\cdot\Bigl[f_{\rm comb}^{{\calA}_1}\cdot {\cal A}_1(\mrec) \, +\cr 
& + &\, (1 - f_{\rm comb}^{{\cal A}_1})\cdot {\cal A}_2(\mrec)\Bigr],
\end{array}
\label{eq:mrecBkgPdf}
\end{equation}
\begin{equation}
\begin{array}{lll}
\M_{\qqbar}(\mrec) & = & f_{\qqbar}^{\rm erf} \cdot {\rm erf}(\mrec-m_{{\rm ep}}); \sigma_{\qqbar})\, +\cr
& + &\,(1 -f_{\qqbar}^{\rm erf})\cdot {\cal A}_{1}(\mrec).
\end{array}
\label{eq:mrecBkgPdf2}
\end{equation}
The parameter $m_{{\rm ep}}$ represents simultaneously the two Argus endpoints 
and the error function inflection point.

The Fisher discriminant PDF $\F_i$ 
is parameterized by two Gaussian functions for each event type $i=(\BB,\qqbar)$,
having standard deviations $\sigma^L_i$ and $\sigma^R_i$, and common mean $\mu_i$:
\begin{equation}
\hspace*{-5pt}\F_i(\fisher) \propto\biggl\{ 
\begin{matrix}
\exp\left[-(\fisher - \mu_i)^2 / 2(\sigma^L_i)^2\right] &\fisher < \mu_i \Bt \cr
\Tp \exp\left[-(\fisher - \mu_i)^2 / 2(\sigma^R_i)^2\right] &   \fisher > \mu_i.
\end{matrix} \\[0.2cm] 
\end{equation}
Since the Fisher variable is designed to discriminate between \qqbar\ and \BB\ events, 
we expect the Fisher discriminant for signal events to be indistinguishable from that of \BB\ combinatorial events. 
We have verified this expectation with \mc\ studies, and thus use the same Fisher discriminant to describe both event types.

\subsubsection{\dt\ \pdf's}
\label{sec:signalpdf}
The $\dt$-dependent part of the \pdf\ is a convolution of the form
\begin{equation}
\begin{array}{ll}
\T'_i(\dt, \dtErr, \stag) =&\\ 
\int d\dttrue\; \T_i(\dttrue, \stag)\,\R_i(\dt - \dttrue, \dtErr),&
\end{array}
\label{eq:sig-pdf-conv}
\end{equation}
where $\T$ is the distribution of $\dttrue$, the true decay time difference,
and $\R$ is a resolution function that parameterizes detector resolution 
and systematic offsets in the measured positions of vertices. 

Taking into account the mistag probability and the effect of tags due to
the unreconstructed \Dz, the \dttrue\ signal \pdf\ in
Eq.\,\ref{eq:sig-pdf-conv} can be written as
\bea
\begin{array}{lll}
\T_{\rm sig} & = & {1\over 4 \tau_b} ~ e^{-{|\dttrue|}/{\tau_b}}\cdot\Bigl\{
 1-\stag\,\Delta\omega(1-\alpha) +\cr
&& + \,\stag\,(1-2\omega)\,(1-\alpha)  \\ 
&& \cdot [C \cos(\dm{d}\dttrue) + \,S\,\sin(\dm{d}\dttrue)] \Bigr\},
\end{array}
\label{eq:sig-pdf-final}
\eea
where the time-dependent \CP asymmetry parameters $S$ and $C$ are the object of the measurement discussed in the present article 
and $\alpha$ (see Sec.\,\ref{sec:mistag}) is the fraction of events in which the tagging track is from the unreconstructed \Dz.
We parameterize possible detector effects leading to a small difference between the mistag probability of $\Bz$ tags  ($\omega^+$)
and that of $\Bzb$ tags ($\omega^-$), by using the average mistag rate $\omega\equiv (\omega^+ + \omega^-)/2$ 
and the mistag rate difference $\Delta \omega\equiv\omega^+ - \omega^-$ as parameters of the \pdf.

Since the \BB\ combinatorial background is dominated by non-\CP\ final states, the \CP\ asymmetry
is expected to be negligible. However, we allow the \pdf\ to accommodate some contamination from
\CP final states. Therefore, we parameterize the \BB\ background \dttrue\ distribution with a \pdf\ similar to that for
signal events given in Eq.\,\ref{eq:sig-pdf-final}. We also add a fraction $f_{\delta}$ of a $\delta$-function, 
to allow for a zero-lifetime component:
\begin{equation} 
\begin{array}{ll}
\T_{\rm comb} = & f_{\rm comb}^{\delta}\cdot\delta(|\dttrue|)\,\Bigl(1-\stag\,\Delta\omega_{\rm comb}^{\delta}\Bigr )\, + \cr
&+(1-f_{\rm comb}^{\delta})\cdot {1\over 4 \tau_{\rm comb}} ~ e^{-{|\dttrue|}/{\tau_{\rm comb}}}\cr
&\cdot\Bigl\{1-\stag\,\Delta\omega_{\rm comb} +\cr 
&+\,\stag\,\cdot \Bigl [\Cbkg \cos(\dm{d}\dttrue) +\cr 
&+ \,\Sbkg\,\sin(\dm{d}\dttrue)\Bigr ] \Bigr\}.
\end{array}
\label{eq:BBbkg-pdf-final}
\end{equation}
The second term of the \pdf\ is obtained from Eq.\,\ref{eq:sig-pdf-final} with
$\omega=\alpha=0$, as these are not defined for background events. The $C_{\rm
comb},S_{\rm comb}$ parameters describe small fluctuations
in the \dttrue\ distribution of background events and possible \CP\ event contamination, leading to a small effective \CP\ violation value.

The $\Delta\omega$ parameters, which for signal events
is the difference in the mistag probabilities for \Bz\ and \Bzb, allow for
differences in the number of events tagged as a \Bz\ or \Bzb\ in
the same background sample.
We use this \pdf\ to describe both the $\Bz\Bzb$ and $\Bp\Bm$ components.

The \pdf\ for the background due to continuum events is modeled with a simple exponential
decay distribution plus a fraction $f_{\delta}$ of a $\delta$-function:
\bea
\begin{array}{lll}
\T_{\qqbar} &= &  f_{\qqbar}^{\delta}\cdot\Bigl(1-\stag\,\Delta\omega_{\qqbar}^{\delta}\Bigr)\cdot\delta(|\dttrue|)\, + \\
& + & (1-f_{\qqbar}^{\delta})\cdot\Bigl(1-\stag\Delta\omega_{\qqbar}\Bigr)\cdot {1\over 4 \tau_{\qqbar}} ~ e^{-{|\dttrue|}/{\tau_{\qqbar}}}
\end{array}
\label{eq:cont-pdf-final}
\eea
where the parameters $\Delta\omega_{\qqbar}^{\delta}$ and
$\Delta\omega_{\qqbar}$ allow for
differences in the number of events tagged as a \Bz or \Bzb in this sample.

\subsubsection{Resolution Functions}
\label{sec:res}
The functions $\T'_i$ of the measured time difference \dt, to be used in the fits, are obtained by convolving the $\T_i$ \pdf's of 
Eq.\,\ref{eq:sig-pdf-final},\,\ref{eq:BBbkg-pdf-final},\,\ref{eq:cont-pdf-final} 
with the  appropriate resolution function for events of type $i$ ($i$=sig, comb, \qqbar).

The resolution functions are parameterized as the sum of
three Gaussian functions:
\bea 
\R_i(t_r, \dtErr) &=& 
        f^n_i\, \G^n_i(t_r, \dtErr) \\
	&+& (1 - f^n_i - f^o_i)\, \G^w_i(t_r, \dtErr) \nonumber\\ 
        &+& f^o_i\, \G^o_i(t_r),\nonumber
\label{eq:res}
\eea
where $t_r = \dt - \dttrue$ is the residual of the $\dt$
measurement, and $\G^n_i$, $\G^w_i$, and $\G^o_i$ are the ``narrow'', ``wide'', and ``outlier'' Gaussian functions. 
The narrow and wide Gaussian functions incorporate information from the $\dt$
uncertainty $\dtErr$, and account for systematic offsets in the
estimation of $\dtErr$ and the $\dt$ measurement. They have the form
\begin{eqnarray}
\label{eq:gRes}
\G^k_i(t_r, \dtErr) &\equiv& 
        {1 \over \sqrt{2\pi} \, s^k_i\, \dtErr}\\
&&  \cdot\exp\left(-\,{\left(t_r - b^k_i\dtErr\right)^2  
        \over 2 (s^k_i\, \dtErr)^2}\right),\nonumber 
\end{eqnarray}
where the index $k$ takes the values $k=n,w$ for the narrow and wide
Gaussian funcions, and $b^k_i$ and $s^k_i$ are parameters determined by
fits.
The outlier Gaussian function, describing a small fraction of events
with badly measured $\dt$, has the form
\begin{equation}
\G^o_i(t_r) \equiv 
        {1 \over \sqrt{2\pi} \, s^o_i  }
  \exp\left(-\,{\left(t_r - b^o_i \right)^2  
        \over 2 (s^o_i)^2}\right).
\label{eq:gOut}
\end{equation}
In all fits, the values of $b^o_i$ and $s^o_i$ are
fixed to 0~ps and 8~\ps, respectively, and are later varied to
evaluate systematic uncertainties.

\subsection{Analysis Procedure}
\label{sec:analProc} 
After the event selection described in Sec.\,\ref{sec:cuts} is complete, the rest of the analysis 
proceeds with a series of unbinned maximum-likelihood fits, performed simultaneously on the on- and
off-resonance data samples and independently for the lepton-tagged and kaon-tagged events.
 The procedure can be logically divided in the following three steps, which we shall discuss in detail in the following paragraphs:
\begin{enumerate}

\item In the first step we determine the signal fraction $f_{\rm sig}$ in Eq.\,\ref{eq:main-pdf} and the shape of
$\M(\mrec)$ and $\F(F)$  in Eq.\,\ref{eq:prod-pdf} for the different classes of events
(signal and backgrounds, kaon and lepton tagging categories). This is done by fitting data with the 
\pdf 
\begin{equation}
\P_i(\mrec, F)  =\M_i(\mrec) \, \F_i(F)
\label{eq:kin-pdf},
\end{equation}
ignoring the time dependence; we refer to this step as the kinematic fit.

\item In the second step we determine the tagging dilution due to wrong tag assignments.

\item In the last step we perform the time-dependent fit to the data. 
We fix all parameter values obtained in the previous steps and use the full \pdf\ of Eq.\,\ref{eq:prod-pdf} 
to determine the parameters of the resolution functions, $\T'_i(\dt,
\dtErr, \stag)$, and the \CP\ asymmetry values $C$, $S$ of the signal and of the
\BB\ combinatorial background component.
\end{enumerate}
The fitting procedure has been validated using both full \mc\ and, where the requested
number of events would be too large, the technique of ``toy'' \mc. In a toy \mc, events are described by a small number 
of variables which are generated according to our \pdf's.

\section{Results}
\label{sec:results}
%
%
Event selection yields
the numbers of events listed in the top two rows of Table~\ref{tab:eventYield}.
\begin{table}[htbp] 
\caption{Event selection yield. The first uncertainty shown is statistical, while the second uncertainty on the number of continuum events 
accounts for a 1\% relative uncertainty on the on-peak and off-peak luminosities.}
\begin{center}
\begin{tabular*}{0.45\textwidth}[t]{r@{\extracolsep{\fill}}rr@{\extracolsep{\fill}}} \hline\hline
  				&  \multicolumn{2}{c}{\Tp\# of events} 		\\ 
			&  \multicolumn{1}{c}{kaon tag} &  \multicolumn{1}{c}{lepton tag}\Bt \\\hline
\Tp on-peak     	& $61179$  				&$20855$\\ 
off-peak  		 	& $1025  $  				&$51$\\ 
continuum  	 	& $9814\pm 307\pm 196$  	&$488\pm 68\pm 10$\\ 
\BB\Bt				& $51365\pm 364$			&$20367\pm 69$\\ \hline
\Tp$N_{{\rm sig}}$\Bt 	& $3843 \pm 397$   		&	$1129 \pm 218$	\\\hline\hline
\end{tabular*}
\end{center}
\label{tab:eventYield}
\end{table}
The third and fourth rows show the number of continuum and \BB\ events calculated, using
Eq.\,\ref{eq:fbb}, from the number of off-peak events in the second row. 
The numbers of signal events in the last line of the table are calculated using the signal
fractions obtained from the kinematic fit described in the next section.
\subsection{Kinematic Fit}
\label{sec:kinFitProc}
We begin by fitting the shape of our signal, $\M_{{\rm sig}}(\mrec)$, using a large
sample of  \mc\ signal events. The parameters most relevant to determine directly
the signal fraction in the data, and consequently our final result for $S$ and $C$, 
will be released again in the final kinematic fit. They are (refer to Eq.\,\ref{eq:mrecSigPdf}): 
the Gaussian fraction $f_G$, mean value $m_G$, and standard deviation $\sigma_G$,
and are shown in the last section of Table~\ref{tab:kinFit}.

Next we fit the Fisher $\F_{\qqbar}$ and recoil mass
$\M_{\qqbar}$ distribution to the off-peak data sample. As the number of
off-resonance events selected in the lepton tagged sample is too small to
yield convergence, we set the lepton tag sample parameters to the corresponding values obtained from
the fit to the kaon tag sample.
Due to the small continuum fraction in the lepton sample, we judge that
this does not introduce any significant systematic effect.
The $\F_{\qqbar}$ parameters are fixed in all subsequent fits. 

We initialize the parameters of the \BB\ combinatorial background \pdf\ directly from the
data, using a sample of events in which the contribution of signal events is much reduced.
We obtain this sample by combining a \Dstar\ with a pion of wrong sign charge (WS sample). 
We have verified, both on \mc\ and in the $\mrec$ sideband for data, (1.836--1.856\gevcc), 
that the shape of the $\M(\mrec)$ distribution for combinatorial \BB background
is well described by that of the WS data sample.

To evaluate a possible contribution from a peaking component in the \BB\ background events, we have allowed the Gaussian fraction $f_G$ in Eq.\,\ref{eq:mrecSigPdf}
to float in a fit to a sample of \mc background events; this fraction is found to be $0.000\pm 0.002$, and is therefore set to zero.

Finally we fit the on-peak data sample, leaving as free parameters 
the fraction $f_{{\rm sig}}$ of signal events in the \BB\ component, 
some of the shape parameters of the continuum 
and \BB\ combinatorial background $\M_{{\rm comb}}$, 
some of the signal parameters in $\M_{{\rm sig}}$, and the shape parameters of the Fisher discriminant $\F_{\BB}$.
Table~\ref{tab:kinFit} summarizes the results and provides information about which parameters are released in the
fit (statistical uncertainties given) and which ones are taken from previous fits
(no uncertainty given).

\begin{figure}[!htbp] 
  \begin{center}
    \includegraphics[width=0.49\textwidth]{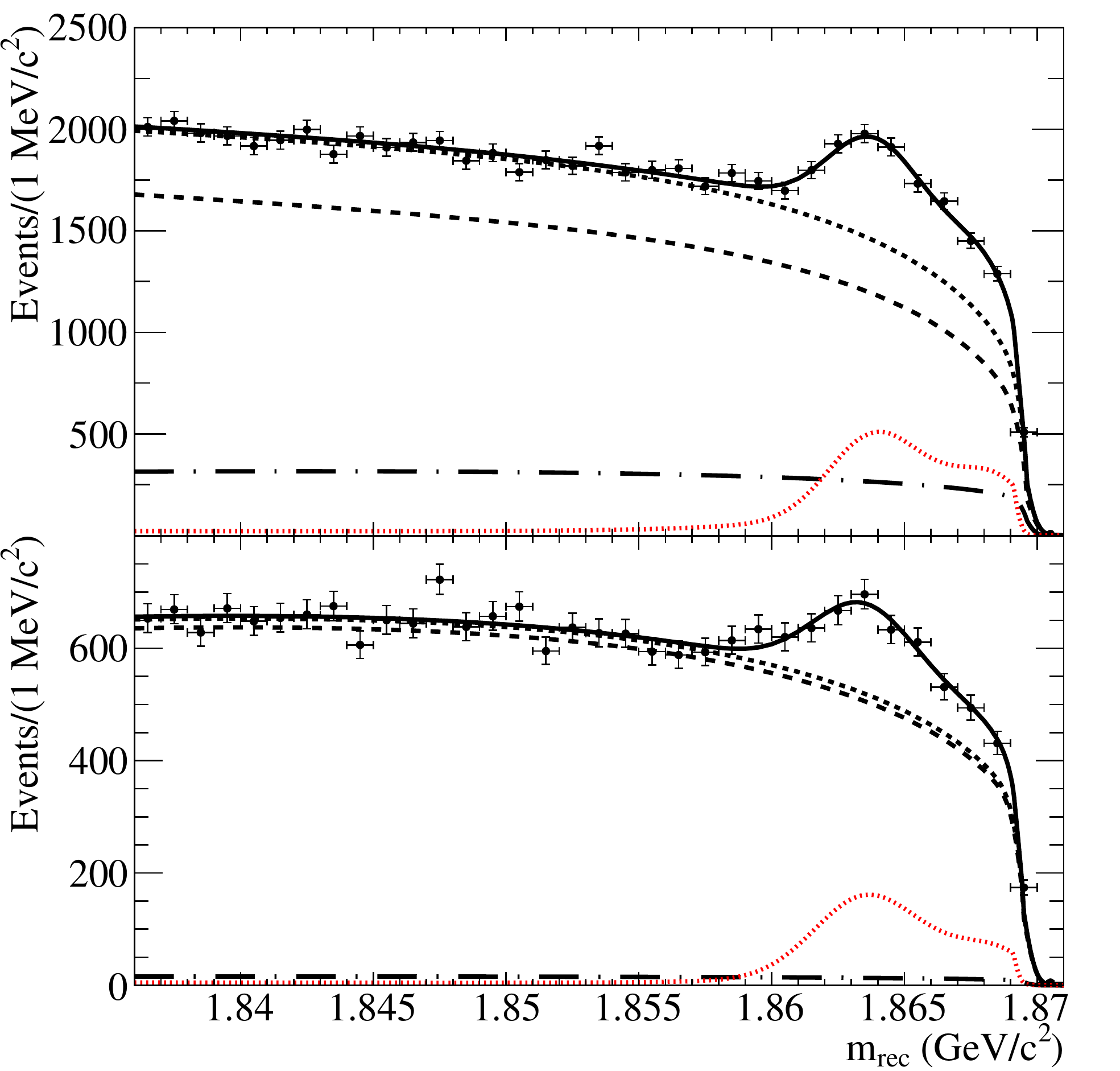}
    \vspace{-15pt}
    \caption{Result of the kinematical fit of kaon (top) and lepton (bottom) tagged data events, with \pdf's overlaid:
 	total \pdf\ (solid line), total background (\BB\ + continuum, short-dashed line), \BB combinatorial
	background (dashed line), continuum  $u,d,s,c$ background
(dot-dashed line) and signal (dotted red line).}
\label{fig:mrecKL}
\end{center}
\end{figure}

\begin{figure}[!htbp] 
  \begin{center}
    \includegraphics[width=0.49\textwidth]{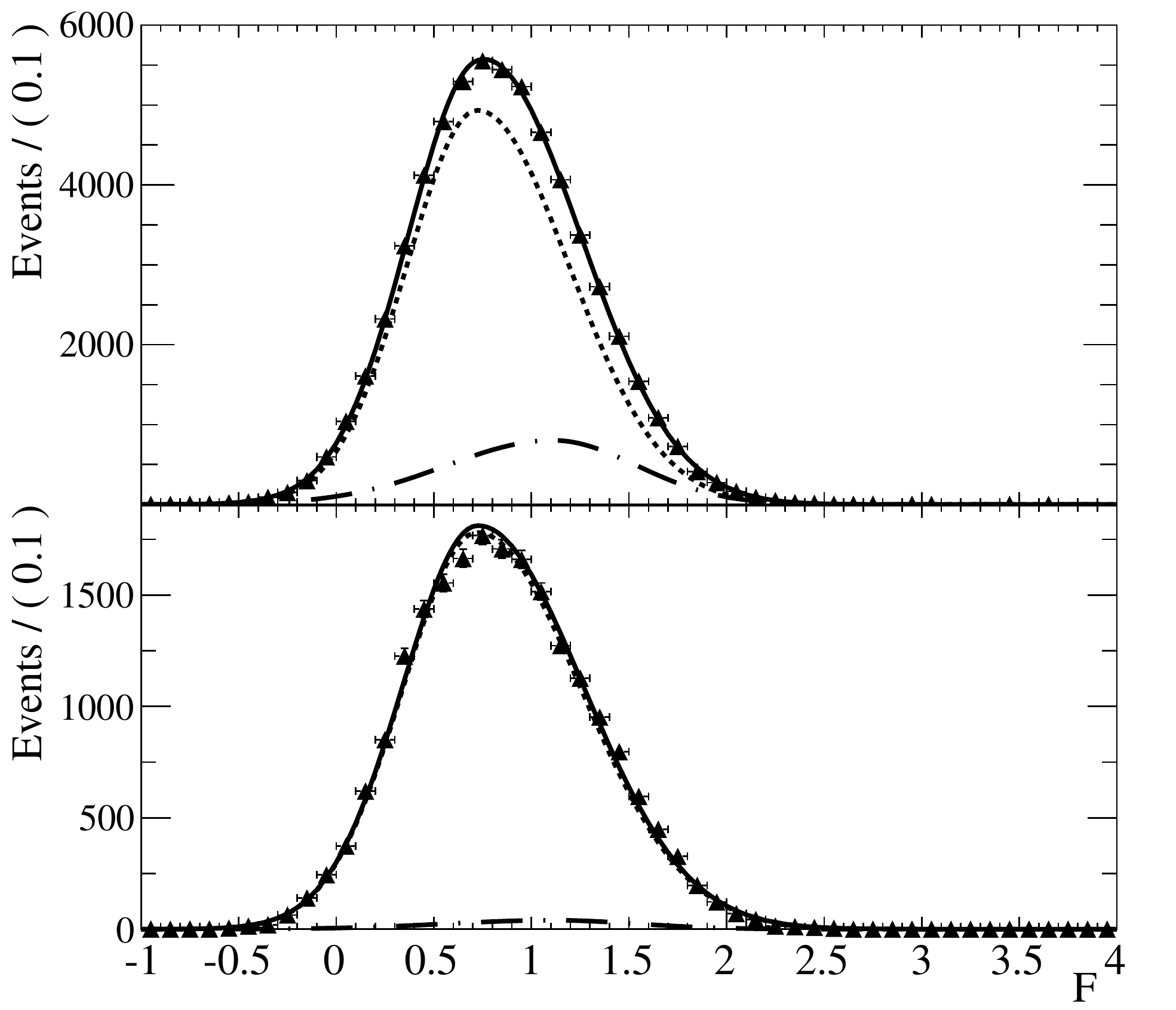}
    \vspace{-15pt}
    \caption{Result of the kinematical fit of kaon (top) and lepton (bottom) tagged events for the Fisher function, with \pdf's overlaid:
 	total \BB\ (solid line), \BB\ (dashed line) and continuum $u,d,s,c$ background (dash-dotted line).}
\label{fig:fishKL}
\end{center}
\end{figure}

The final results of the kinematic fits for the kaon and lepton tagged sample 
are shown in Fig.\,\ref{fig:mrecKL} and Fig.\,\ref{fig:fishKL}.
\begin{table*}[tbp] 
\caption{Results of the final kinematic fits. The values of fixed parameters are given without uncertainties.}
\begin{center}
\begin{tabular*}{0.95\textwidth}[t]{c@{\extracolsep{\fill}}r@{\extracolsep{\fill}}l@{\extracolsep{\fill}}r@{\extracolsep{\fill}}r} \hline\hline
\Tp \pdf	                & parameter	        &  description\Bt         &    kaon tags                &   lepton tags    	         \\\hline
\hfil                           &\Tp $f_{{\rm sig}}$\Bt   & Signal fraction         &  $ (7.5 \pm 0.7 ) \% $	  &  $ (5.5 \pm 1.1 ) \%$      \\\cline{2-5} 
\multirow{3}{*}{$\F_{\BB}(F)$}	&\Tp $\mu_{\BB}$ \Bt  	&                         & $0.723 \pm 0.005 $          &  $0.721 \pm 0.009$	         \\ 
				&$\sigma_{\BB}^L$  \Bt	&                         & $0.361 \pm 0.003 $          &  $0.380 \pm 0.006$	          \\ 	
				&$\sigma_{\BB}^R$\Bt  	&                         & $0.469 \pm 0.004 $          &  $0.532 \pm 0.006$	          \\ \hline
\multirow{5}{*}{$\M_{\qqbar}(\mrec)$}&\Tp $f^{A}$        & Argus fraction         & $ 1.0$		          &  $ 1.0$	                  \\ 
	                            &$m_{\rm ep}$  	& Argus endpoint (\gevcc)   & $1.8696$		        &  $1.8696$	                  \\
	                            &$c$ 		& Argus exponent          & $    -17 \pm 9 $	        &  $   -17$ 	                  \\ 
	                            &$f_{\rm comb}$ 	& erf fraction   & $    0.52 \pm 0.15$         &  $  0.52$	                  \\ 
 	                            &$\sigma_{\rm comb}$\Bt  & erf width (\gevcc)	  & $  0.0005 \pm 0.0002$       &  $0.0005$	                  \\ \hline
\multirow{6}{*}{$\M_{{\rm comb}}(\mrec)$}&\Tp $f^{{\cal A}_1}$ & Argus fraction	  & $  0.27 \pm 0.06 $	        &  $  1.0 $	                  \\ 
	                               &$m_{\rm ep}$     & Argus end point (\gevcc)        & $1.8696		$	 &  $1.8695$ 		          \\
	                               &$c_1$	        & ${\cal A}_1$ exponent   & $ -49 \pm 38 	$	&  $-15 \pm 2 $	          \\ 
	                               &$c_2$	        & ${\cal A}_2$ exponent   & $ -0.56 \pm 0.25$	        & 		--	          \\ 	
                                       &$f_{\qqbar}$ 	& erf fraction      & $ 0.26 \pm 0.04$	        &  $  0.41\pm 0.06$             \\ 
	                               &$\sigma_{\qqbar}$\Bt & erf width (\gevcc)     & $(75 \pm 9)\cdot 10^{-5}$ &  $(72\pm 2)\cdot 10^{-5}$      \\\hline
\multirow{3}{*}{$\M_{{\rm sig}}(\mrec)$}&\Tp $f_{G}$ 	      & Gaussian fraction & $  0.46 \pm 0.06 $          &  $ 0.64\pm 0.12$	          \\ 
	                              &$m_G$  		      & Gaussian peak (\gevcc)    & $  1.8638 \pm 0.0002 $      &  $ 1.8635\pm 0.0003$          \\ 
	                              &$\sigma_G$\Bt	      & Gaussian width (\gevcc)	  & $  0.0017 \pm 0.0002 $      &  $ 0.0019\pm 0.0003$          \\ 
\hline\hline
\end{tabular*}
\end{center}
\label{tab:kinFit}
\end{table*}

\subsection{Determination of mistag probabilities}
\label{sec:mistag}

A common problem of analyses using the partial reconstruction technique is
that a fraction of the tracks used in tagging may belong to
the unreconstructed \Dz, leading to a mistag of the event. 
As the tracks originating from the missing \Dz\ tend to align to its
direction of flight, this fraction can be reduced by applying a constraint
on the cosine of the CM opening angle $\theta_{{\rm tag}}$ between the
tagging track and the direction of the unreconstructed \Dz. 
We require $\cos\theta_{{\rm tag}}\leq 0.75~(0.50)$ for the kaon (lepton) tagged sample. 

We find from signal \mc\ that before this requirement 26\% (13\%) of kaon
(lepton) tags originate from a missing \Dz. 
We call these events ``\D-tags'', while ``$T$-tags'' are those in which the tagging
track (either direct or cascade) is from the tag \B.

To reduce dependency on \mc\ 
the fraction $\alpha$ of \D-tags remaining after the
$\cos\theta_{{\rm tag}}$ constraint is measured using data, as explained below.

Defining the total number $N_T$ ($N_D$) of $T$ ($D$) tags, 
the number $N^l_T$ ($N^l_D$) of them that satisfy the $\cos\theta_{{\rm tag}}$ requirement, 
and the number $N^g_T$ ($N^g_D$) of them that do not, 
$\alpha$ is given by:
\begin{equation}
\alpha = N_D^l/(N^l_D+N_T^l)=\frac{f^Dp^D}{(f^Dp^D+(1-f^D)p^T)},
\label{eq:tagDilAlmostFinal}
\end{equation}
where 
$p^T=N_{T}^{l}/N_{T}$ ($p^D=N_{D}^{l}/N_{D}$) is the probability, 
taken from signal \mc, for a $T$-tag (\D-tag) to be from a track
satisfying the $\cos\theta_{{\rm tag}}$ cut.
The fraction of \D-tags, $f^D=N_D/(N_T+N_D)$, is given by
\begin{equation}
f^D = (p^T-f^l)/(p^T-p^D).
\label{eq:tagDilInvert}
\end{equation}
The fraction $f^l=N^l/(N^l+N^g)$ is obtained from the kinematic fit of the
data: $N^l$ is the number of signal events that have 
$\cos\theta_{{\rm tag}}\leq 0.75~(0.50)$ 
and $N^g$ is the number of signal events
with $\cos\theta_{{\rm tag}}\geq 0.75~(0.50)$ for kaon (lepton) tag events.

In this way we obtain the values $\alpha=0.12\pm 0.04$ for kaon tags and $\alpha=0.00\pm0.02$ for lepton tags,
as shown in Table~\ref{tab:mistag-data}, where we also list the mistag parameters $\omega$
and $\Delta\omega$, $\alpha$, $\tau_b$, and $\dm{d}$ that we will need in the final \dt\ fit. 

We use information from signal \mc\ events to determine the mistag probability $\omega=0.201\pm0.002~(0.101\pm0.002)$
and mistag difference  $\Delta\omega=-0.011\pm0.003~(0.001\pm0.005)$ for the kaon (lepton) tagged samples.
We use the world average value for $\dm{d}$\,\cite{ref:PDG2010}, 
and the values fitted to signal \mc\ for $\tau_b$, as discussed in Sec.\,\ref{sec:cuts}.

\begin{table}[htb] 
\caption{Values of mistag parameters and $\alpha$ used in the final fit. The $b$ lifetime values were obtained from the fit of signal \mc.
The statistical uncertainties are given.}
\begin{center}
\begin{tabular*}{0.45\textwidth}[t]{r@{\extracolsep{\fill}}r@{\extracolsep{\fill}}r} \hline\hline
\Tp parameter\Bt	     		&  kaon tags & lepton tags  \\ \hline
\Tp $\omega$ & $  0.201 \pm 0.002 	$	& $  0.104 \pm 0.002 	$ \\ 
$\Delta \omega$ & $ -0.011 \pm 0.003  $& $  0.001 \pm 0.005 $	 \\ 
$\alpha$ & $  0.12 \pm 0.04  	$& $  0.0 \pm 0.02  	$	 \\ 
$\tau_b\Bt ~{\rm (ps)} $  & $ 1.458  \pm 0.014	$& $ 1.518  \pm 0.018	$	 \\ \hline
$\Tp \dm{d}~{\rm (ps^{-1})}$\Bt  & \multicolumn{2}{c}{$ 0.507  \pm 0.004 	$}\\ \hline\hline
\end{tabular*}
\end{center}
\label{tab:mistag-data}
\end{table}
 
\subsection{Time Dependent Fit}
\label{sec:dtFitProc}
After the kinematic fit is complete and mistag probabilities are determined, 
we can proceed to the final \dt\ fit to extract the \CP-violating parameters $S$ and $C$.

We start by fitting the signal \dt\ \pdf\ and its resolution function using a pure signal \mc sample; the parameters determined in this way 
will be used to initialize the signal \pdf, and some of them will be left free again in the final \dt\ fit. 

Next we fit the resolution function parameters and the effective lifetime of the
continuum background, using the off-peak data sample. 
For the kaon tag sample, the data strongly disfavor a component with nonzero lifetime, 
therefore we fix $f_\delta=1$, and leave free in the final \dt\ fit only $\Delta\omega_\delta$ from Eq.\,\ref{eq:cont-pdf-final}.
For lepton tags, as the real data sample is too small, we obtain resolution and \dt\ parameters from continuum \mc.

We use the continuum parameters obtained above in the next fit stage, where
we obtain the \BB\ background resolution function and \dt\ parameters using real data,
by restricting the fit to events in a sideband region (1.836--1.856~\gevcc) of the \Dz\ recoil
mass distribution. According to \mc studies the fraction of signal
events in this sideband is negligible, and we set it to zero.
We fit simultaneously the resolution and lifetime parameters, shown in sections 1 and 3 of Table~\ref{tab:final-data}.
The fitted values of $C_{{\rm comb}}$ and $S_{{\rm comb}}$ are consistent with 0 as expected.

We are now in a position to perform the final \deltat\ fit, in which we release parameters from the signal, 
continuum and \BB\ combinatorial background \deltat\ and resolution models, as detailed in Table~\ref{tab:final-data},
always with the convention that parameters free (fixed) in the final \dt\ fit are shown with (without) a fit uncertainty. 

\begin{figure}[!htbp] 
  \begin{center}
   \includegraphics[width=0.49\textwidth]{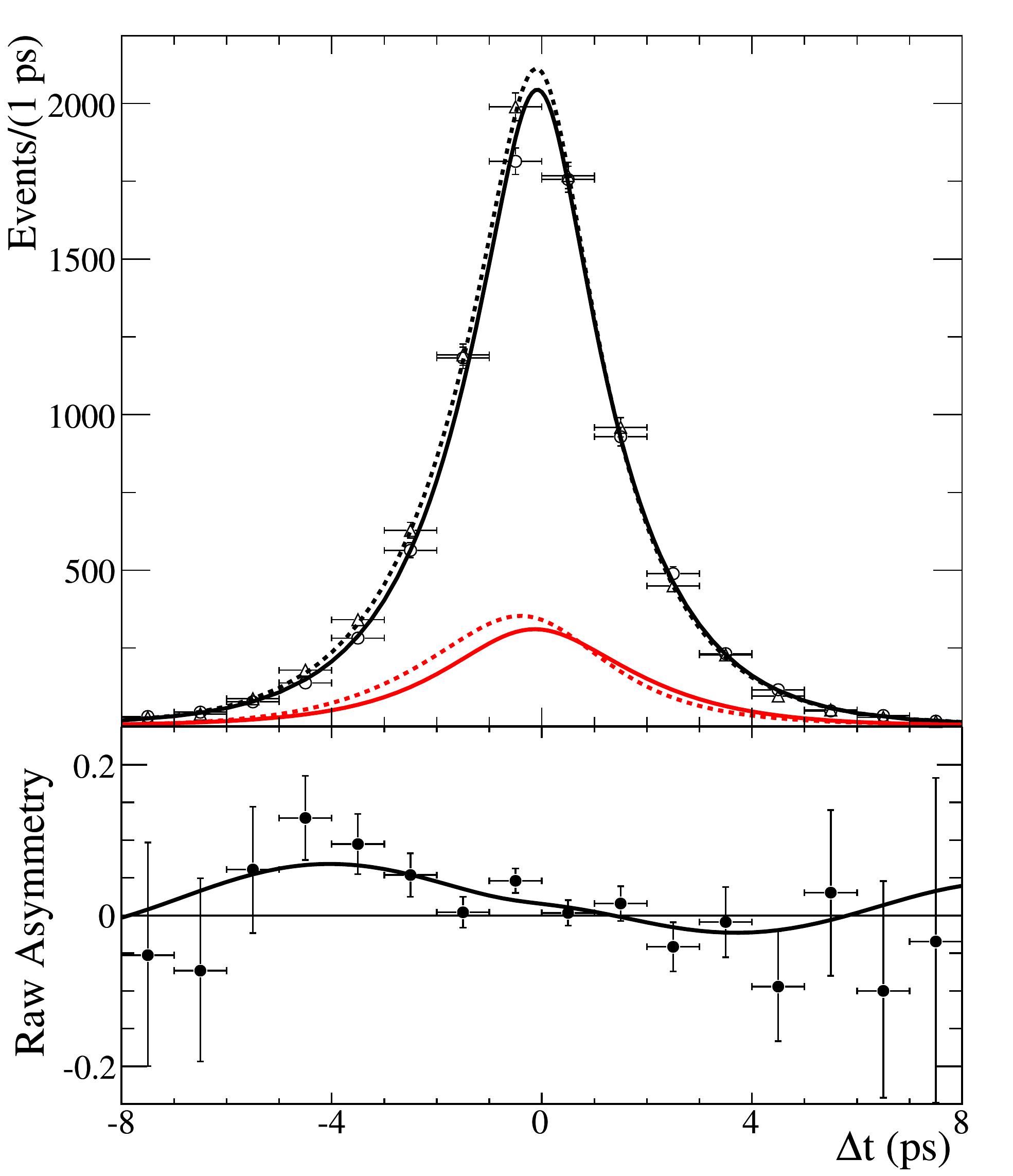}
    \vspace{-15pt}
    \caption{Top: \dt\ distribution for \Bz (dashed) and
\Bzb (solid) kaon tags; the lower curves are the corresponding signal \pdf s. 
Bottom: raw time-dependent \CP asymmetry.
Only data in the restricted signal region $\mrec >1.860~\gevcc$ are shown.} 
    \label{fig:final-k-asy-SR}
  \end{center}
\end{figure}

\begin{figure}[!htbp] 
  \begin{center}
    \includegraphics[width=0.49\textwidth]{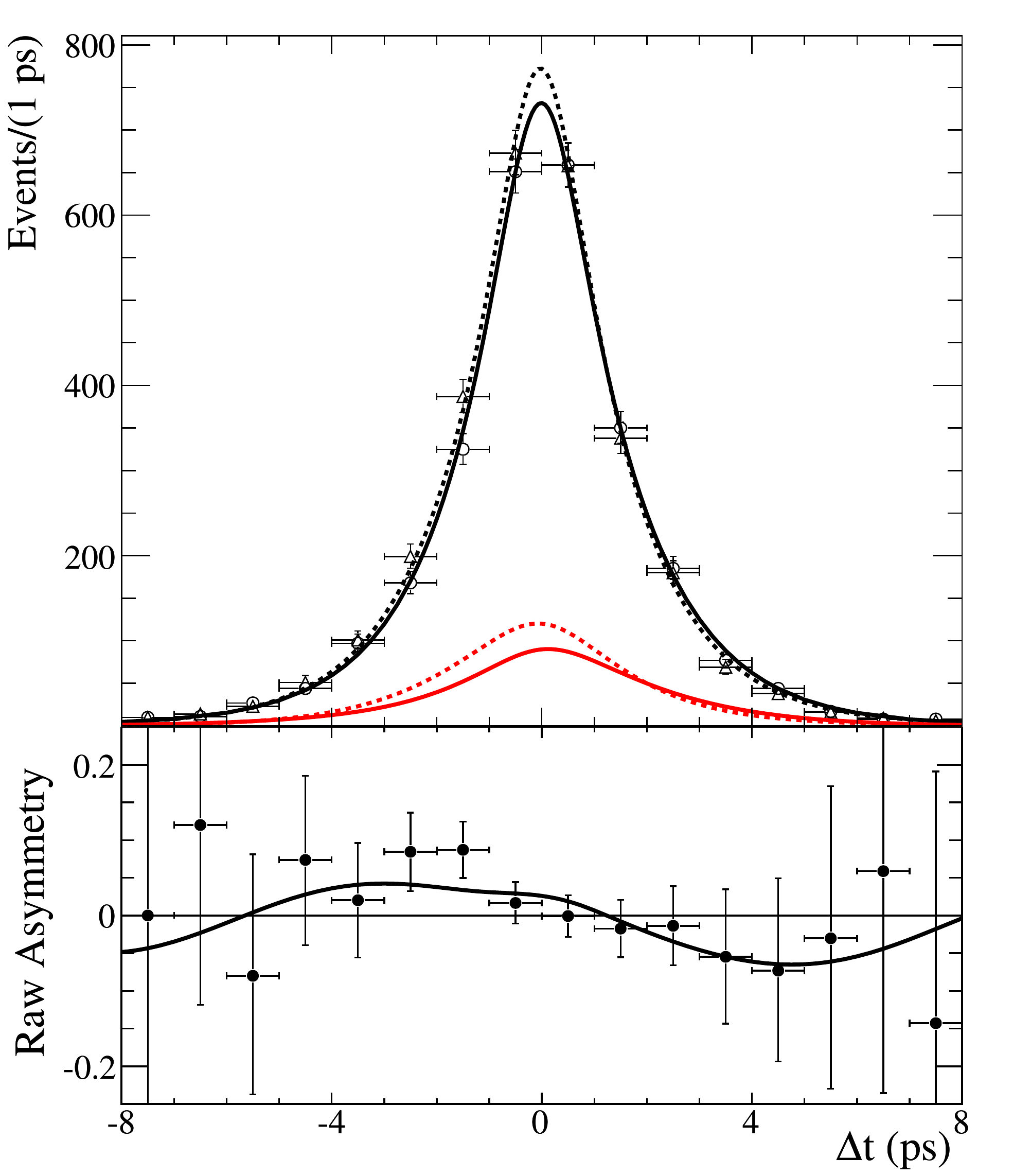}
    \vspace{-15pt}
    \caption{Top: \dt\ distribution for \Bz (dashed) and
\Bzb (solid) lepton tags; the lower curves are the corresponding signal \pdf s. 
Bottom: raw time-dependent \CP asymmetry. 
Only data in the restricted signal region $\mrec >1.860~\gevcc$ are shown.} 
    \label{fig:final-l-asy-SR}
  \end{center}
\end{figure}

The results are also shown in Fig.\,\ref{fig:final-k-asy-SR} (\ref{fig:final-l-asy-SR}) for the kaon (lepton) tagged samples, where we plot
the \dt\ distributions separately for \Bz\ and \Bzb\ tags, together with the time-dependent raw \CP asymmetry
\begin{equation}
A(\dt)=\frac{N_{S_{{\rm tag}}=1}(\dt)-N_{S_{{\rm tag}}=-1}(\dt)}{N_{S_{{\rm tag}}=1}(\dt)+N_{S_{{\rm tag}}=-1}(\dt)}.
\end{equation}
For presentation purposes, only data in the restricted signal region $\mrec >1.860~\gevcc$
are shown in Fig.~\ref{fig:final-k-asy-SR} and Fig.~\ref{fig:final-l-asy-SR}, in order to amplify signal/background ratio
and be able to see the oscillation. The signal fractions in this region become
$\approx 24\%$ and $\approx 18\%$ for kaon and lepton tags respectively.

This requirement is not applied to the fit sample, so our numeric results apply to the whole
signal region $\mrec >1.836~\gevcc$.
\begin{table*}[htb] 
\caption{Result of the final full fit. The values of fixed parameters are given without uncertainties.}
\begin{center}
\begin{tabular*}{0.95\textwidth}[t]{c@{\extracolsep{\fill}}l@{\extracolsep{\fill}}l@{\extracolsep{\fill}}r@{\extracolsep{\fill}}r} \hline\hline
		\pdf						&	\Tp parameter\Bt &  description                            &kaon tags & lepton tags  \\ \hline
\multirow{8}{*}{\BB\ resolution model} 	&\Tp $b^n$&  offset of narrow Gaussian   		        & $-0.16 \pm 0.01	$     	&  $ -0.022\pm 0.014$\\
								&$b^o$	 & offset of outlier Gaussian (ps)		& $0.0			$      &$ 0.0$ \\ 
								&$b^w$	 &  offset of wide Gaussian   			        & $-1.0 \pm 0.2$     	&$ -0.7\pm 0.7$ \\ 
								&$f^n$    &  fraction of narrow Gaussian 		            	& $ 0.93 \pm 0.01	$     	&$  0.977\pm 0.004$ \\ 
								&$f^o$    &  fraction of outlier Gaussian		            	& $ 0.008 \pm 0.001$ 	&$  0.006\pm 0.002$ \\ 
								&$s^n$    &  see eq.\,\ref{eq:gRes}  		            	& $ 1.03 \pm 0.03	$     	&$  1.02\pm 0.02$ \\ 
								&$s^o$    &  see eq.\,\ref{eq:gOut}   (ps)		        & $8.0			$      &$ 8.0$ \\ 
								&$s^w$\Bt &  see eq.\,\ref{eq:gRes}  		        & $3.0			$      &$ 5.6$ \\ \hline
continuum \deltat 			&\Tp $\Delta\omega_{\qqbar}^{\delta}$\Bt & see eq.\,\ref{eq:cont-pdf-final}& $-0.04 \pm 0.02	$     	&$0.03$\\ \hline
\multirow{6}{*}{\BB\deltat} 	&\Tp $f_{\rm comb}^{\delta}$ 	& see eq.\,\ref{eq:BBbkg-pdf-final} 	    	& $ 0.10 \pm 0.02	$     	&$  0.25\pm 0.02$\\ 
						&$\Delta\omega_{\rm comb}^{\delta}$ & see eq.\,\ref{eq:BBbkg-pdf-final} 	 & $ 0.04 \pm 0.12	$     	&$ -0.08\pm 0.07$\\ 
						&$\Delta\omega_{{\rm comb}}$ 	& see eq.\,\ref{eq:BBbkg-pdf-final} & $-0.025 \pm 0.012$	&$  0.012\pm 0.021$\\ 
						&$\tau_{{\rm comb}}$    	& effective lifetime  (ps)    		    			& $ 1.318 \pm 0.023$ 	& $  1.272\pm 0.031$\\
						&$C_{{\rm comb}}$       	& cosine coefficient       		    			& $-0.022 \pm 0.024$   &$ -0.024\pm 0.041$\\ 
						&$S_{{\rm comb}}$\Bt       	& sine coefficient     			    	& $ 0.004 \pm 0.014$	&$ -0.023\pm 0.024$\\ \hline
\multirow{6}{*}{signal resolution model} &\Tp $b^n$&  offset of narrow Gaussian   			& $-0.35 \pm 0.09	$     	&$ -0.3\pm 0.2$  \\
								&$b^w$	 &  offset of wide Gaussian   			        & $ 8 \pm 3		$     	& -- \\ 
								&$f^n$    &  fraction of narrow Gaussian 		            	& $ 0.992 \pm 0.007$	&$  1.0\pm 0.1$ \\ 
								&$f^o$    &  fraction of outlier Gaussian		            	& $0.0			$     & --\\ 
								&$s^n$    &  see eq.\,\ref{eq:gRes}  		            	& $ 1.13 \pm 0.12	$     	& $  1.17\pm 0.21$\\ 
								&$s^w$\Bt &  see eq.\,\ref{eq:gRes}  		        & $2.6			$     &-- \\ \hline\hline
\multirow{2}{*}{signal $\deltat$}	&\Tp $C$   	& 	                         			&$ +0.117 \pm 0.111$	& $ +0.195\pm 0.147$\\
									&	$S$\Bt 	& 	                         			&$-0.417 \pm 0.159$	& $ -0.210\pm 0.200$\\\hline\hline
\end{tabular*}
\end{center}
\label{tab:final-data}
\end{table*}

\section{Systematic Uncertainties}
\label{sec:SystErr}
%
%
Our systematic uncertainties on the \CP-violating parameters $S$ and $C$ are
summarized in Table~\ref{tab:syserror}. We discuss here the most significant ones.

Most systematic uncertainties in Table~\ref{tab:syserror} 
are due to imperfect knowledge of one single parameter fixed in the final \dt\ fit, 
having little or no correlation with uncertainties of other parameters. They have been treated by varying 
them by $\pm1\sigma$ and repeating the final \dt\ fit leaving 
only the parameters $S$ and $C$ free to vary.

Uncertainties in the first two lines have a different character because they are due
to parameter sets, in which correlations among parameters belonging to one set are non-trivial. 
Given the low signal-to-background ratio, 
correct modelling of the background shape and signal fraction in the kinematic fit is crucial, 
especially because the $\M(\mrec)$ and $\F(F)$ \pdf's parameters are fixed in the final \deltat fit. 
Consequently, we devised a procedure to evaluate the associated systematic uncertainties,
that would also preserve the correlations among parameters belonging to a set.
\begin{table}[htb]
\caption{\label{tab:syserror} Systematic uncertainties evaluated for $C$
and $S$. Uncertainties in the top section are independent for kaon
and lepton tags, those in the bottom section are correlated.}
 \begin{center}
  \begin{tabular}{lrrrr}\hline\hline 
        \Tp                             & \multicolumn{2}{c}{kaon tags} & \multicolumn{2}{c}{lepton tags}       \\ 
    Systematic source\Bt                        &  $C$          	& $S$           	&     $C$       	& $S$           \\ \hline
    \Tp Kinematic fit parameters                &  $0.013$      	& $0.034$       	& $0.023$       	& $0.057$       \\
    Continuum $\Delta t$ fit parameters         &  $0.002$      	& $0.001$		& $-$       		& $-$       \\
    Signal $s_w$ 				&  $0.0002$      	& $0.0007$		& $-$       		& $-$       \\
    \BB\ combinatorial $s_w$ 			&  $0.017$      	& $0.0007$		& $0.001$       	& $0.005$       \\
    Signal tag side ($\omega$)     & $0.012$       	& $0.045$       	& $0.002$       	& $0.002$       \\
    Mistag difference ($\Delta \omega$)              & $0.007$       	& $0.0004$      	& $0.007$       	& $0.0009$      \\
    Signal \CP side (\alphaDz)& $0.006$       	& $0.017$       	& $0.002$       	& $0.002$       \\
    Peaking background                          & $0.0002$      	& $0.0003$      	& $0.0002$      	& $0.00004$     \\
    Fit bias (MC statistics)                    & $0.011$       	& $0.018$      		& $0.012$       	& $0.019$       \\
    Tag interference from DCSD       	& $0.030$      		& $0.002$  		& --    		& --           \\ 
    \Bt\Bz lifetime variation                      & $0.0002$      	& $0.002$       	& $0.0003$      	& $0.004$       \\\hline
    \Tp\dm{d} variation                      & $0.0003$      	& $0.001$       	& $0.0004$      	& $0.002$       \\
    SVT misalignment                            & $0.003$       	& $0.007$       	& $0.002$       	& $0.004$       \\
    Boost uncertainty\Bt                           & $0.002$       	& $0.006$       	& $0.005$		& $0.007$	\\\hline
    \Tp Total\Bt   		& $0.042$	& $0.062$	& $0.028$	& $0.061$ \\\hline\hline
  \end{tabular}
 \end{center}
\end{table}

For each set of parameters in $\M$ or $\F$ that become fixed at any stage of our fits,
and are not released again in the final \dt\ fit,
a large number $N_t$ of toy \mc experiments of the same size as the data are generated and fitted,
and the values of parameters in the $N_t$ experiments are saved. 
Evaluation of the systematics due to a set of parameters in subsequent fits
(in which they become fixed) is made by repeating the latter fits many times over,
using the same event sample, but fixing parameters in the set to different values
every time, taken from one of the $N_t$ experiments. 
In this way, we propagate the variation associated to parameter sets
from one fit to the next one, and preserve correct correlations among them.
We applied this procedure to obtain the uncertainties in lines 1 and 2 of Table~\ref{tab:syserror},
which for lepton tagged events are the main source of systematics.

For lepton tags we find that only one Gaussian is sufficient to describe
the resolution function ($f^n=1$). The systematic due to the signal $s^w$ was evaluated only for kaon tagged events.

Since the mistag parameter $\omega$ is obtained from
\mc with a very small statistical uncertainty (see Table~\ref{tab:mistag-data}),
we verified the agreement between \mc\ and data using a control sample of
self-tagging $\btodstpipm$ events. As a result of this study, we assign a very
conservative uncertainty of $15\%$ on $\omega$, and evaluate the 
associated systematic by repeating the final \dt\ fit 
varying its central value by $\pm15\%$. 
This is the largest systematic uncertainty for kaon tagged events.

We estimate the systematic uncertainty associated with fixing the peaking background fraction $f_G$ in Eq.\,\ref{eq:mrecSigPdf}
to zero by setting it to $\pm 0.002$ for both the kaon and lepton tag samples, repeating the fit, 
and taking the largest deviation from the value fitted with $f_G=0$ as the systematic uncertainty.

The signal \mrec spectra for the \CP-even and \CP-odd components are different, the latter being slightly harder.
This may cause a small acceptance difference of our event reconstruction and selection, leading
 to a systematic shift in the $C$ and $S$ measurement.  
We have carefully evaluated this effect and found it to be negligible.

As the \dstpdstm\ final state is a superposition of
\CP-even and \CP-odd wave functions, the measured values $S$ and $C$ from 
our data only represent a weighted average of these
components, with their inverse squared errors as weights. 
Since the background shape is not uniform as it goes to zero at the
kinematical limit, the weight of the \CP-odd component could be enhanced
with respect to the \CP-even one by the lower background level in the high
mass region.
To evaluate this effect, we perform \deltat fits in the two extreme \mc
configurations, adding to the background sample a pure
\CP-odd ($\Rperp=1$) or \CP-even ($\Rperp=0$) sample of signal events, respectively. 
The number of signal events in both cases is equal to the number of signal events found in data. 
We find that the differences in the errors of $S$ and $C$ are negligible in these two cases and we do not
assign a systematic uncertainty to this effect.

As discussed in Sec.\,\ref{sec:dtFitProc}, $\tau_b$ and $\deltamd$ are fixed to the values listed in Table~\ref{tab:mistag-data}.
We assign the systematic uncertainty due to these
assumptions by varying their nominal values of $\pm 1\sigma$,
and taking half the difference in the fitted values of $C$ and $S$ so
obtained. 

To evaluate bias on $C$ and $S$ in our fit, we apply the fit procedure to pure signal \mc\ events 
and compare the results for $C$ and $S$ to the generated ones; no significant bias has been observed in either.
We therefore quote the statistical uncertainty on these \mc\ measurements
as the associated systematic uncertainties.

To measure the systematic uncertainty related to imperfect knowledge of the time
measurement due to uncertainty in the boost or possible uncorrected
misalignment of the SVT, we repeat the time-dependent
fit with different sets of realistic misalignments of the SVT and \dt\ scaling
factors. The maximum observed shift with respect to the nominal fit is quoted as the uncertainty.

\newcommand{\Cfit}      {\ensuremath{C_\mathrm{fit}}\xspace}
\newcommand{\Sfit}       {\ensuremath{S_\mathrm{fit}}\xspace}
\newcommand{\Czero}   {\ensuremath{C_0}\xspace}
\newcommand{\Szero}   {\ensuremath{S_0}\xspace}
\newcommand{\dprime} {\ensuremath{\delta^\prime}\xspace}
\newcommand{\calK}     {\ensuremath{\cal K}\xspace}

An important source of systematic uncertainty in our analysis is represented 
by interference effects from doubly Cabibbo-suppressed decay amplitudes 
on the tagging side of the event.
The non-leptonic \B-meson decays used for tagging are
dominated by amplitudes containing a $b\to c\bar u d$ transition,
thus ensuring the correlation of the tagging particle (typically a
kaon) with the flavor of the originating $b$ quark. However,
$\bar b\to \bar u c \bar d$ transitions could also contribute,
although they are suppressed\cite{ref:tagsidecp} by a factor
$r' \simeq |(V_{ub}^* V_{cd})/(V_{cb}V_{ud}^*)| = 0.02$.

As discussed in detail in Ref.\,\cite{ref:tagsidecp}, this effect
cannot be simply reabsorbed into the mistag probability $\omega$ because
the allowed and doubly Cabibbo-suppressed amplitudes can interfere,
and thus effectively change the \deltat probability density function.

Since our \deltat PDF assumes $\rprime=0$ and therefore does not
include these effects, the $C$, $S$ parameters measured by our fit
will be different from the observables without tag-side interference
by a calculable quantity. 

To evaluate the systematic effect in our measurement due to neglecting
small terms in the PDF with $\rprime\ne0$, we follow the prescription
in Ref.\,\cite{ref:tagsidecp} and perform a simple toy \mc of
$\delta C\equiv\Cfit-\Czero$ and $\delta S\equiv\Sfit-\Szero$, finding the results reported in Table~\ref{tab:syserror}.
The lepton tags are not affected
by this issue.

\section{Physics Results}
\label{sec:physRes}
%
%
The final results for $C$ and $S$, with their correlation coefficient $\rho$,
including only the statistical uncertainty for kaon and lepton tags, are:
\begin{eqnarray}
\begin{array}{lll}
C & = +0.12\pm 0.11& \quad\multirow{2}{*}{$\rho=0.0601,\quad$ kaon tags,}\\ 
S & = -0.42\pm 0.16&\nonumber
\end{array}
\label{eq:result-k}
\end{eqnarray}
\begin{eqnarray}
\begin{array}{lll}
C & = +0.20\pm 0.15& \quad\multirow{2}{*}{$\rho=0.0730,\quad$ lepton tags.}\\ 
S & = -0.21\pm 0.20\nonumber
\end{array}
\label{eq:result-l}
\end{eqnarray}
The two samples are statistically independent of each other and can therefore be combined;
their statistical uncertainties can be combined in quadrature, but the systematic ones need a more careful treatment. 

Indeed, several of the systematic effects listed in Table\,\ref{tab:syserror} are independent for the kaons and lepton
tags and are combined in quadrature, while the others are combined taking into account their correlation.
Finally we get the combined results of this analysis of
\begin{eqnarray}
\begin{array}{lll}
C & = &  +0.15\pm 0.09\pm 0.04 \quad\multirow{2}{*}{$\rho=0.0649$.}\\ 
S & = &  -0.34\pm 0.12\pm 0.05\nonumber
\end{array}
\label{eq:result-combined}
\end{eqnarray}

\subsection{Extraction of \Sp and \Cp}
The measured values of $S$ and $C$ that we
obtain from data only represent a weighted average of the
\CP even and \CP odd wave function components. 
If penguin amplitudes can be neglected then $S_+=-S_-$, $C_+=-C_-$  
and the value of the \CP-even components \Sp and \Cp, 
which we are interested in, can be obtained using the relations:
\begin{eqnarray}
\begin{array}{lll}
C&=&C_+\\
S&=&S_+ \left(1-2 R_\perp\right),\nonumber
\end{array}
\label{eqn:S_C_and_Rperp-bis}
\end{eqnarray}
where the factor $(1-2R_\perp)$ represents the dilution introduced by
the \CP-odd component $R_\perp$ in the signal. 
To compute $S_+$
 we use the value measured by \babar\ of ($R_\perp=0.158\pm 0.029$)\,\cite{Aubert:2009rx}, 
where the uncertainty is the combined statistical and systematic.
To evaluate the related systematic uncertainty, we vary this value by $\pm 1\sigma$. 
We obtain
\begin{eqnarray}
\begin{array}{lll}
\Cp & = & +0.15  \pm 0.09 \pm 0.04 \\
\Sp & = & -0.49 \pm 0.18 \pm 0.07 \pm 0.04, \nonumber
\end{array}
\label{eq:SpCp}
\end{eqnarray}
where the uncertainties shown are statistical and systematic; the third uncertainty is
the contribution from the error on $R_\perp$ described above.

\section{Summary}
\label{sec:summary}
%
%
We have measured the time-dependent \CP asymmetry parameters $C$ and $S$ in \btodstdst decays, from which 
we have extracted the \CP-even components \Sp\ and \Cp.
This result is an independent determination of the \CP-violating parameters
of $b\rightarrow(c\cbar)d$ transitions and is compatible with previous
measurements from \babar\,\cite{Aubert:2009rx} and
Belle\,\cite{Vervink:2009sy} using fully reconstructed decays. 
It also agrees well with the Standard Model expectation of negligible
contributions to the decay amplitude from penguin diagrams and thence with $\Sp=-\sin
2\beta$.

\begin{acknowledgments}
\label{sec:ack}
We are grateful for the 
extraordinary contributions of our \pep2\ colleagues in
achieving the excellent luminosity and machine conditions
that have made this work possible.
The success of this project also relies critically on the 
expertise and dedication of the computing organizations that 
support \babar.
The collaborating institutions wish to thank 
SLAC for its support and the kind hospitality extended to them. 
This work is supported by the
US Department of Energy
and National Science Foundation, the
Natural Sciences and Engineering Research Council (Canada),
the Commissariat \`a l'Energie Atomique and
Institut National de Physique Nucl\'eaire et de Physique des Particules
(France), the
Bundesministerium f\"ur Bildung und Forschung and
Deutsche Forschungsgemeinschaft
(Germany), the
Istituto Nazionale di Fisica Nucleare (Italy),
the Foundation for Fundamental Research on Matter (The Netherlands),
the Research Council of Norway, the
Ministry of Education and Science of the Russian Federation, 
Ministerio de Ciencia e Innovaci\'on (Spain), and the
Science and Technology Facilities Council (United Kingdom).
Individuals have received support from 
the Marie-Curie IEF program (European Union) and the A. P. Sloan Foundation (USA).

\end{acknowledgments}


\begin{thebibliography}
\bibitem{cconjimplied}Throughout this article, charge conjugate decay modes are implied.
\end{thebibliography}
\end{document}